\documentclass[fleqn,usenatbib]{mnras}

\usepackage{newtxtext,newtxmath}

\usepackage[T1]{fontenc}
\usepackage{ae,aecompl}
\usepackage{autobreak}

\usepackage{graphicx}	
\usepackage{amsmath}	
\usepackage{amssymb}	
\usepackage{color}

\title[The origin of warm absorbers]{Thermally driven wind as the origin of warm absorbers in AGN}

\author[M.~Mizumoto et al.]{
Misaki Mizumoto$^{1}$\thanks{E-mail: misaki.mizumoto@durham.ac.uk, mizumoto.misaki@gmail.com (MM)},
Chris Done$^{1,3}$,
Ryota Tomaru$^{2,3}$,
\& Isaac Edwards$^{1}$
\\
$^{1}$Centre for Extragalactic Astronomy, Department of Physics, University of Durham, South Road, Durham DH1 3LE, UK\\
$^{2}$Department of Physics, The University of Tokyo, 7-3-1 Hongo, Bunkyo, Tokyo 113-0033, Japan \\
$^{3}$Kavli Institute for the Physics and Mathematics of the Universe (WPI), University of Tokyo, Kashiwa 277-8583, Japan\\
}

\date{Accepted XXX. Received YYY; in original form ZZZ}

\pubyear{2019}

\begin{document}
\label{firstpage}
\pagerange{\pageref{firstpage}--\pageref{lastpage}}
\maketitle

\begin{abstract}
Warm absorbers are present in many Active Galactic Nuclei (AGN), seen
as mildly ionised gas outflowing with velocities of a few hundred to a
few thousand kilometres per second. These slow velocities imply a
large launch radius, pointing to the broad line region and/or torus as
the origin of this material. Thermal driving was originally suggested
as a plausible mechanism for launching this material but recent work
has focused instead on magnetic winds, unifying these slow, mildly
ionised winds with the more highly ionised ultra-fast outflows. 
Here we use the recently developed quantitative models for
thermal winds in black hole binary systems to predict the column
density, velocity and ionisation state from AGN. Thermal winds are
sensitive to the spectral energy distribution (SED), so we use
realistic models for SEDs which change as a function of mass and mass
accretion rate, becoming X-ray weaker (and hence more disc dominated)
at higher Eddington ratio. These models allow us to predict the launch
radius, velocity, column density and ionisation state of thermal winds
as well as the mass loss rate and energetics. While these match well to some of the
observed properties of warm absorbers, the data point to the presence of additional wind material, 
most likely from dust driving. 
\end{abstract}

\begin{keywords}
galaxies: nuclei -- X-rays: galaxies -- X-rays: ISM --  quasars: general
\end{keywords}



\section{Introduction}

X-ray observations of active galactic nuclei (AGNs) often reveal the
presence of mildly ionised material, with multiple absorption lines
from partially ionised oxygen, neon, and iron in the 0.5--2~keV
bandpass. These lines are blueshifted, indicating outflow
velocities of a few hundred to a few thousand km s$^{-1}$. These `warm absorbers' are best
studied with high resolution X-ray data from gratings, the Reflection Grating Spectrometer (RGS) on
board {\it XMM-Newton} (e.g.~\citealt{sak01} and the compilations of
\citealt{lah14}) and the High/Low Energy Transmission Grating (HETG/LETG) on {\it Chandra} (e.g.\ \citealt{kas00b,kaa00}
and the compilation of \citealt{mck07}). The inferred mass outflow
rate is often comparable to or even larger than the mass accretion
rate onto the supermassive black hole, so this must impact on the
available material for accretion \citep{blu05,lah14}. 
However, its kinetic energy is rather small,
generally less than 1\% of bolometric luminosity
\citep{blu05,lah14}. Thus the warm absorbers are probably not
important in setting $M$-$\sigma$ relation as these typically require a
wind with kinetic power of 0.5--5\% of the bolometric luminosity \citep{hop10}.

There are three main models for producing the warm absorber outflows:
radiation pressure, magnetic force, and thermal pressure.  The
radiation force overcomes gravity when $L > L_{\rm Edd}$, where 
$L_{\rm Edd}$ is the Eddington limit, which is defined from Thompson scattering
on free electrons. Most AGNs with warm absorbers are
sub-Eddington, so continuum radiation pressure cannot be the main
mechanism. However, there can be other processes which enhance the coupling
of the gas to the radiation field. An additional cross-section, $\sigma_i$,  leads to a decrease in
the luminosity at which a wind can be driven, to $L_{\rm Edd}/ (1+ M)$ where 
$M=\sigma_i/\sigma_{\rm T}$ is the force multiplier. The force multiplier can be very large for low ionisation
gas due to the enormous numbers of ultraviolet (UV) line (bound-bound) as well as edge (bound-free) 
transitions, allowing radiation pressure to drive strong winds in sub-Eddington AGNs
(e.g.\ \citealt{pro04}). However, this is
most efficient where there is strong UV radiation, which is the inner disc in 
most bright AGN. Typically, winds have terminal velocity which is of order the 
escape velocity from their launch point so these winds are much faster than those observed
in warm absorbers \citep{blu05}.

Winds can also be driven by centrifugal acceleration along magnetic
field lines anchored in the disc \citep{bla82,kon94,fuk10}.  However,
these winds depend on the (currently unknown) magnetic field
configuration, so are impossible to calculate {\it ab initio}.

The third wind launch mechanism is thermal driving
\citep{beg83,woo96}, which was first applied to the warm absorbers by
\citet{kro95} based on numerical studies by \citet{bal93}.
X-rays from the AGN heats any illuminated material up
to the Compton temperature, $T_{\rm IC}$. This is determined only by the
spectrum of the radiation, as photons with $h\nu\ll kT_{\rm IC}$ will
Compton cool the material, whereas photons with $h\nu\gg kT_{\rm IC}$ heat
it. The heated skin expands due to the pressure gradient, producing a
thermally driven wind at radii where the sound speed exceeds the local
escape velocity.

In this paper we focus on the thermally driven wind model as these can
be rather well predicted given the spectral energy distribution (SED)
and luminosity of the source. We use the approach of \citet{don18}, who
recast the analytic thermal wind solutions of \citet{beg83} into a
more tractable form, and used them to show that this is most likely
the origin of the narrow, highly ionised, blueshifted absorption seen
in black hole binaries (see also \citealt{hig18,tom19}). Here we apply this instead to AGN, predicting the thermal wind
using the SED models of \citet{kub18} to track how the Compton
temperature varies with mass and $L/L_{\rm Edd}$. We compare these
predictions with the data to assess the viability of a thermally
driven wind model for the origin of warm absorbers in AGN.

\section{Thermal winds in AGN}
\begin{figure}
	\includegraphics[width=0.8\columnwidth, angle=270]{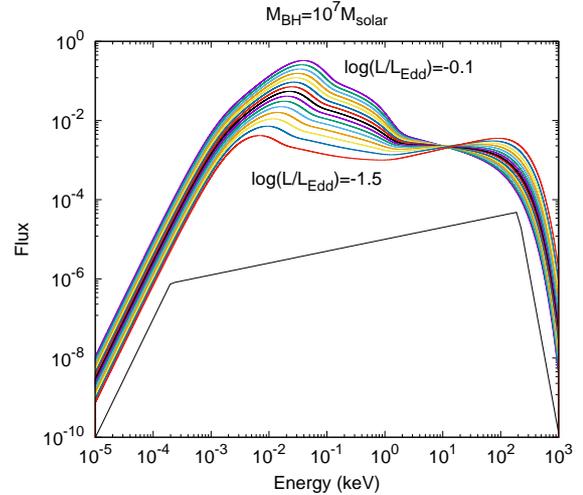}
    \caption{AGN SED model for $M_{\rm BH}=10^7 M_\odot$. 
    The vertical axis has an arbitrary unit of $\nu F_\nu$.
    The colour lines are for Eddington ratios of $\log (L/L_{\rm Edd})=-1.5, -1.4, \dots, -0.1$, from bottom to top at 0.1~keV. 
    The spectra become softer for larger Eddington ratios in this range. 
    We use the same spectral shape for $\log (L/L_{\rm Edd})\leq-1.7$ (black line), which normalisation is changed for the different luminosities.}
    \label{fig:SED}
\end{figure}

\begin{figure}
	\includegraphics[width=1.1\columnwidth, angle=270]{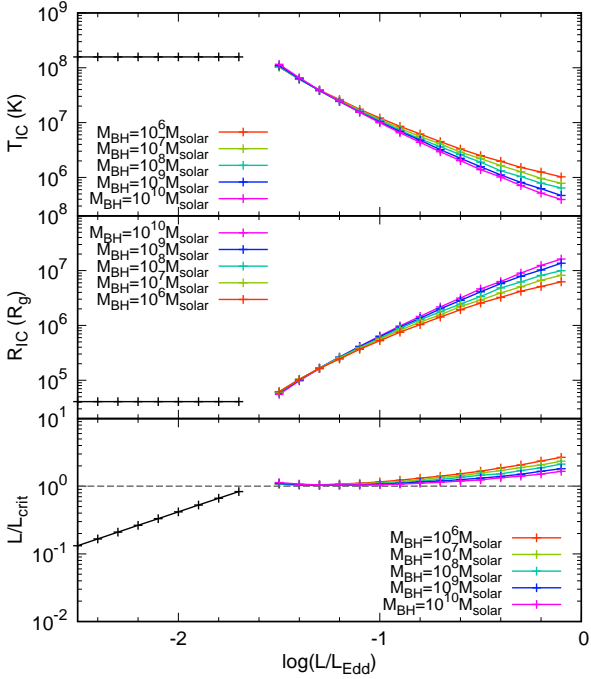}
    \caption{Compton temperature (upper), Compton radius (medium), and critical luminosity (low) for different Eddington ratios and black hole masses. 
    In the low Eddington ratio regime ($\log(L/L_{\rm Edd})<-1.7$) no difference is seen among different black hole masses (black lines).}\label{fig:param}
\end{figure}

Irradiation by X-rays from the inner region heats material to the Compton temperature, defined by 
\begin{equation}
kT_{\rm IC}=\frac{1}{4}\frac{\int E L(E) dE}{\int L(E) dE},
\end{equation}
where $L(E)$ is luminosity.  $T_{\rm
  IC}$ depends on only the shape of SED; harder spectra produce higher $T_{\rm IC}$.  The heated
material expands with the sound velocity of $c_{\rm
  IC}=\sqrt{kT_{\rm IC}/\mu}$, where $\mu=0.63\,m_p$ is the mean
particle mass of ions and electrons for solar abundances.  The Compton
radius ($R_{\rm IC}$) is where this local sound
speed exceeds the escape velocity, i.e., $R_{\rm IC}=GM_{\rm
  BH}/c_{\rm IC}=6.4\times10^4 T_{{\rm IC},8}^{-1} R_g$, where
$T_{{\rm IC},8}=10^{-8}T_{\rm IC}$ and $R_g=GM/c^2$ is the gravitational
radius.

The spectral shape is critical to the thermal wind properties.
Multiple papers have shown that the AGN SED changes systematically as
a function of mass and mass accretion rate 
\citep{vas07,vas09,jin12,don12}. 
We use the specific model {\sc qsosed} of \citet{kub18} to model them. 
  This captures the main trends seen in the data by 
  assuming that the accretion flow forms three different regions, an outer
standard disc where the emission thermalises to the local blackbody
temperature, an intermediate region where the accretion power is
dissipated higher up in the photosphere, producing a `soft X-ray
excess' warm Comptonisation region, and an inner hot flow which is
assumed to have constant hard X-ray luminosity, $L_{\rm X}=0.02L_{\rm
  Edd}$. At low luminosity, almost all of the accretion power is taken
by the hot flow, whereas for $L\sim L_{\rm Edd}$, this forms only a very
small fraction of the bolometric luminosity, as required (see
\citealt{kub18} for details).  Fig.~\ref{fig:SED} shows examples of
the assumed SED for a black hole of mass $10^7M_\odot$ and zero spin
for $L/L_{\rm Edd}=0.03$ to $1$. For $L\le 0.03L_{\rm Edd}$ the
spectrum is assumed to be purely a power law with photon spectral index of
$\Gamma=1.7$, i.e., that the entire disc is replaced by an advection
dominated accretion flow (ADAF).

Fig.~\ref{fig:param} shows the resulting Compton temperature,
$T_{\rm IC}$, and corresponding Compton radius, $R_{\rm IC}$, at which material at
this temperature is able to escape the black hole gravity. The softer
SED at higher $L/L_{\rm Edd}$ gives a lower Compton temperature, so the
radius at which the heated material can escape is larger.  We show
these predictions for a range of black hole masses, from
$10^6M_\odot$ (red) to $10^{10}M_\odot$ (magenta), but this makes little difference as the Compton
temperature is much more sensitive to higher energy photons
(essentially it is a spectral average of $\nu^2 F_v$ rather than $\nu F_v$).
Thus the expected decrease in $T_{\rm IC}$ due to the lower disc temperature at
higher mass only becomes noticeable at the highest $L/L_{\rm Edd}$, where the 
disk component almost completely dominates the spectrum.

\begin{figure}
	\includegraphics[width=0.8\columnwidth, angle=270]{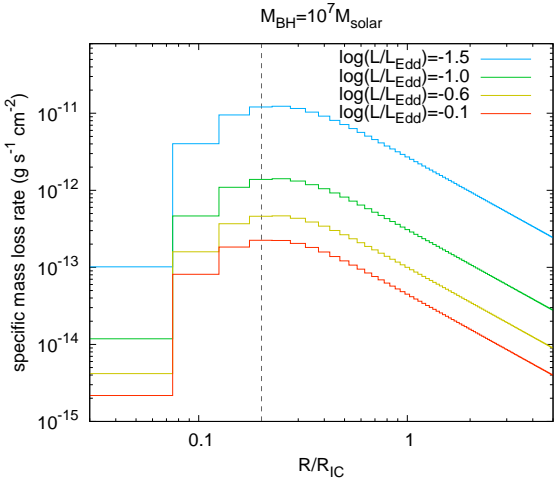}
    \caption{Specific mass loss rate for $M_{\rm BH}=10^7\,M_\odot$. 
    Peaks constantly appear at $R\sim0.2R_{\rm IC}$.
    }\label{fig:specific}
\end{figure}

The luminosity is of secondary importance to the spectral shape as long as it is high enough to 
heat the gas up to the Compton temperature before it escapes. However, if the luminosity is lower
than this critical luminosity, $L_{\rm crit}$, then the gas is heated 
only to a characteristic temperature ($T_{\rm ch} < T_{\rm IC}$).
The critical luminosity and the characteristic temperature are written as 
\begin{equation}
\begin{split}
L_{\rm crit}&=\frac{1}{8}\left(\frac{m_e}{\mu}\right)^{1/2}\left(\frac{m_ec^2}{kT_{\rm IC}}\right)^{1/2}L_{\rm Edd}\\
&\simeq 0.03T_{\rm IC,8}^{-1/2}L_{\rm Edd}.
\end{split}
\end{equation}
and
\begin{equation}
T_{\rm ch}=T_{\rm IC}\left(\frac{L}{L_{\rm crit}}\right)^{2/3}\left(\frac{R}{R_{\rm IC}}\right)^{-2/3}
\end{equation}
\citep{beg83,don18}.
The lower panel of Fig.~\ref{fig:param} shows $L/L_{\rm crit}$ for each of the SEDs.
The luminosity is around or above
the critical luminosity for $\log (L/L_{\rm Edd})\geq-1.5$, predicting that all 
AGN with some UV emitting
outer disc have $L>L_{\rm crit}$ in our models, so can efficiently produce
thermal winds.

\begin{figure}
\centering
	\includegraphics[width=0.8\columnwidth]{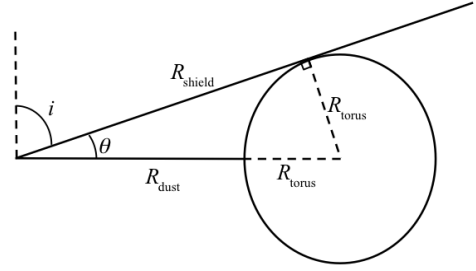}
    \caption{Geometry of the torus and the self-shielding radius. $R_{\rm torus}$ is the radius of the torus, and $R_{\rm dust}$ is the distance from the central black hole to the inner radius of the torus. $R_{\rm shield}$ is the distance from the black hole to the contact point of the torus, beyond which the wind cannot be launched because the seed gas does not exist. $\theta$ is set to be $34^\circ$.
    }\label{fig:geometry}
\end{figure}

\citet{woo96} provided an equation for the specific mass loss rate
(per unit area) at radius $R$ from hydrodynamic simulations as
\begin{align}
  \begin{autobreak}
\dot{m}(R)=\frac{L/c}{4\pi R^2 \Xi_{\rm max} c_{\rm ch}}
\left\{\frac{1+[(0.125L/L_{\rm crit}+0.00382)/\zeta]^2}{1+[(L/L_{\rm crit})^4(1+262\zeta^2)]^{-2}}\right\}^{1/6}
\exp\left\{\frac{-[1-(1+0.25\zeta^{-2})^{-1/2}]^2}{2\zeta}\right\},
  \end{autobreak}\label{eq:specific}
\end{align}
Here the characteristic sound speed is $c_{\rm ch}=\sqrt{kT_{\rm
    ch}/\mu}$, $\zeta=R/R_{\rm IC}$, and the pressure ionisation
parameter marking the base of the X-ray heated atmosphere, $\Xi_{\rm max}$, is assumed constant at $\sim 40$
\citep{don18}.  
This equation matches well with the analytic expectations \citep{beg83,don18} and more
recent hydrodynamic results \citep{hig18}.
Fig.~\ref{fig:specific} shows the results of equation
(\ref{eq:specific}), which are calculated with a step of $0.05R_{\rm
  IC}$.  These peak at $R \sim 0.2R_{IC}$ as long as 
the heating is rapid ($L>L_{\rm crit}$), as is the case here for all AGN 
with $L/L_{\rm Edd}\ge 0.03$. Hence we consider this to be the launch radius for
thermal winds. \citet{don18} showed that in the more general case 
the wind launch radius is 
$R_{\rm in}=0.2R_{\rm IC}$ for $L/L_{\rm crit}>1$ and $0.2R_{\rm
  IC}/(L/L_{\rm crit})$ for $L/L_{\rm crit}\leq1$.

\begin{figure*}
\centering
	\includegraphics[width=0.85\columnwidth, angle=270]{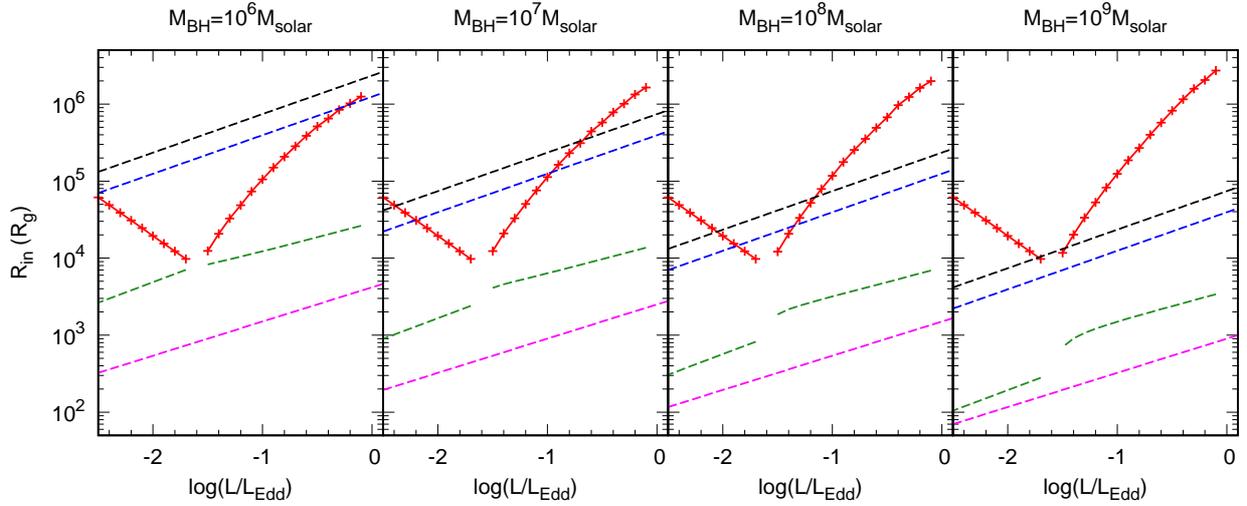}
    \caption{The inner radius of the thermal wind (red) and the characteristic radii of the AGN structure. The dashed black, blue, green, and magenta lines are $R_{\rm shield}$, $R_{\rm dust}$, $R_{\rm BLR}$, and $R_{\rm sg}$, respectively, from top to bottom.
    The $R_{\rm BLR}$ line has a small jump because the assumed SED is changed at this point.}
    \label{fig:param2}
\end{figure*}

\section{Illuminated material}

Thermal winds are produced where the AGN spectrum illuminates cool
material, with typical launch radii of $10^5\,R_g$ for most of the
SEDs considered (Fig.~\ref{fig:param}).  In black hole binaries
it is common for the outer accretion disc to extend to these radii, but 
in AGN self gravity should truncate the disc at a radius \citep{lao89} of
\begin{equation}
R_{\rm sg}=2150M_9^{-2/9}\dot{m}^{4/9}\alpha^{2/9},
\end{equation}
where $M_9=M_{\rm BH}/10^9M_\odot$, $\dot{m}=\dot{M}_{\rm
  acc}/\dot{M}_{\rm Edd}$, and $\alpha$ is the disc viscosity
parameter of \citet{sha73}, assumed here as $\alpha=0.02$
\citep{sta04}. 
Thus thermal winds cannot be produced from AGN discs. 

Nonetheless, there is gas at larger radii, in the self gravitating regime, from either 
the broad line region (BLR) and/or molecular torus \citep{kro01}. 
Both of these are radially extended structures, so we consider that there is a 
continuous distribution of gas connecting the accretion disc to the molecular torus. 
Nonetheless, there are characteristic radii which can be identified within this. 
Reverberation mapping shows that this gas produces the broad H$\beta$ line at
\begin{equation}
\begin{split}
&\log(R_{\rm BLR}/1\,{\rm lt \mathchar"712D day}) =\\
&\qquad\quad(1.527\pm0.031) 
+0.533^{+0.035}_{-0.033} \log (\lambda L_\lambda
/10^{44}\,\mathrm{erg\,s}^{-1})
\end{split}
\end{equation}
\citep{ben13}, 
where $\lambda L_\lambda$ is taken at 5100\AA\ from the AGN SED model (Fig.~\ref{fig:SED}).

The inner radius of the dust torus ($R_{\rm dust}$) is determined by
the dust sublimation temperature, $T_{\rm sub}\simeq1500$~K \citep{bar87}. Dust
evaporates when the irradiation flux reaches $L/(4\pi R^2) = \sigma_{\rm SB}T_{\rm sub}^4$,
where $\sigma_{\rm SB}$ is the Stefan-Boltzmann constant.
Therefore,
\begin{equation}
R_{\rm dust}=\left(\frac{L}{4\pi\sigma_{\rm SB} T_{\rm sub}^4}\right)^{1/2}.
\end{equation}

There should also be an outer radius for illumination of the torus,
depending on the geometry. 
When we simply assume a circular cross-section torus, it has a
self-shielding radius ($R_{\rm shield}$, see
Fig.~\ref{fig:geometry}), as $R_{\rm shield}/R_{\rm
  dust}=\cos\theta/(1-\sin\theta)$, where $\theta$ is the half opening
angle subtended by the torus from the central black hole.  We set
$\theta=34^\circ$ to match the hydrodynamic simulation of
\citet{dor08a}, which assume that the torus obstructs all lines of
light of AGN with inclinations of $i>56^\circ$.  Therefore $R_{\rm
  shield}=1.88R_{\rm dust}$. 
The dust torus may not have a simple toroidal shape, but be conical/flared made by a dusty wind (e.g.~\citealt{eli06,dor12a,cha16}). In this case the dusty gas in the torus can supply more gas to the wind, which will be discussed in section \ref{sec7}.

The BLR and inner and outer torus radii all scale as
$R\propto L^{1/2}\propto (M\dot{m})^{1/2}$.  Hence $R/R_g \propto
(\dot{m}/M)^{1/2}$ so the typical radii of these gas structures
increases with $L/L_{\rm Edd}$ but decreases with increasing black hole mass.
Fig.~\ref{fig:param2} shows how the H$\beta$ broad line radius (green) 
and inner/outer (blue/black) radius of the dusty torus all decrease systematically in terms
of $R_g$ for increasing black hole masses of 
$M=10^6, 10^7, 10^8$ and $10^9M_\odot$, while the
predicted thermal wind launch radius (red crosses) is approximately constant in
$R/R_g$. Thus while the lowest mass AGN can launch a thermal wind from radii 
close to the H$\beta$ emitting part of the BLR, the highest mass AGN 
cannot launch a thermal wind at all as the torus self shields before 
the material can escape in a wind with our assumed geometry. 
The range in wind producing radii is largest when 
$\log(L/L_{\rm Edd})\sim-1.5$, and at lowest black hole masses, so highlighting these system
parameters as the ones where the thermal wind mass loss rates should be largest.

\section{Wind mass, momentum and energy loss rates}

\begin{figure}
\centering
	\includegraphics[width=0.85\columnwidth, angle=270]{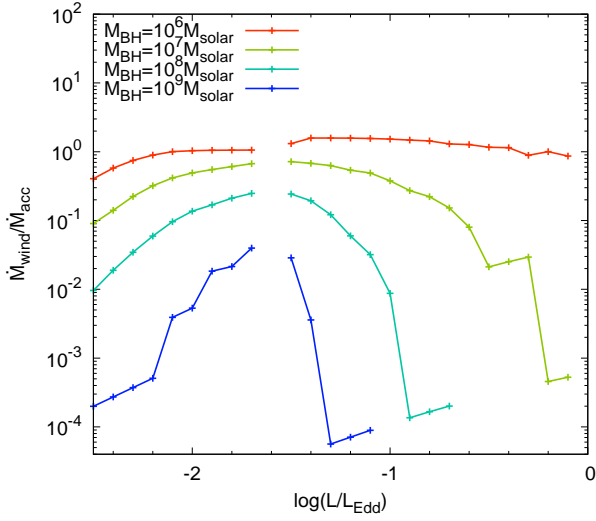}
    \caption{The total mass-loss rate. The vertical axis is normalised by the mass accretion rate.}
    \label{fig:total_massloss}
\end{figure}

The total mass loss rate in the wind is the integration of the
specific mass loss rate up to $R_{\rm shield}$, i.e.,
\begin{equation}
\dot{M}_{\rm wind}=\int_0^{R_{\rm shield}} \dot{m}(R)\times2\times2\pi R dR.
\end{equation}
Fig.~\ref{fig:total_massloss} shows the ratio of the total mass loss
rate to the mass accretion rate ($\dot{M}_{\rm acc}=L/\eta c^2$, where
$\eta=0.057$ for a non-spinning black hole).  The mass-loss is indeed larger
when the black hole mass is smaller and the Eddington ratio is closer to
$10^{-1.5}$ as expected from the discussion above. It exceeds the mass
accretion rate for the most efficient case.

\begin{figure}
\centering
	\includegraphics[width=0.8\columnwidth, angle=270]{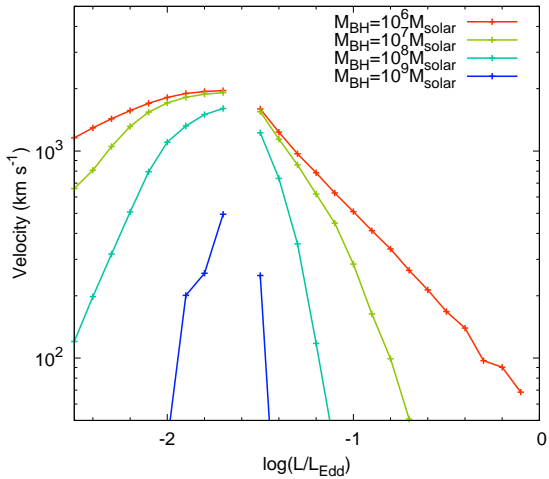}
    \caption{Wind velocity, calculated as the mass loss weighted average sound speed}
    \label{fig:velocity}
\end{figure}

\begin{figure}
\centering
	\includegraphics[width=0.8\columnwidth, angle=270]{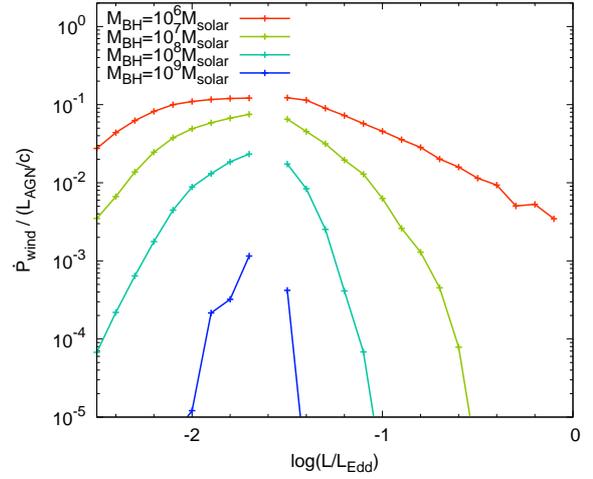}
    \caption{Momentum of the winds. The vertical axis is normalised by $L_{\rm AGN}/c$.}
    \label{fig:momentum}
\end{figure}

\begin{figure}
\centering
	\includegraphics[width=0.8\columnwidth, angle=270]{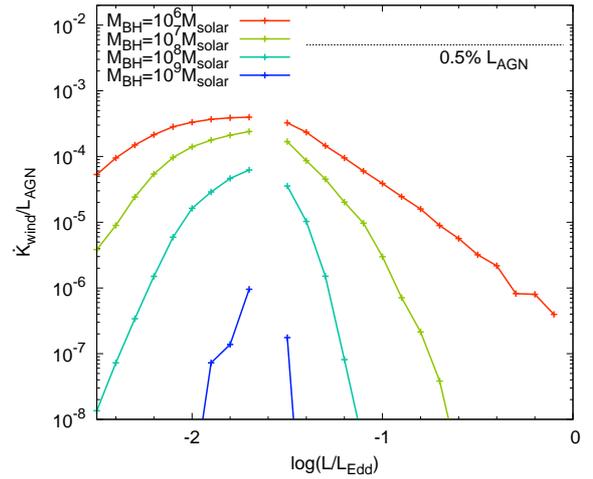}
    \caption{The energy-loss rate. The vertical axis in the upper panel is normalised by the AGN luminosities.}
    \label{fig:energy}
\end{figure}

We calculate the expected wind velocity following \citet{don18} as the
mass loss weighted average sound speed ($c_{\rm ch}$).  The wind
velocity becomes slower for larger Eddington ratios
(Fig.~\ref{fig:velocity}), which due to their lower $T_{\rm IC}$
(Fig.~\ref{fig:param}). This enables us to calculate the 
momentum and kinetic energy carried by the wind, as shown in 
Figs.\ \ref{fig:momentum} and \ref{fig:energy}.  Because the wind
velocity is much slower than the light velocity
($\lesssim1500$~km~s$^{-1}=0.05c$), the momenta are much less than $L_{\rm
  AGN}/c$ and the energy loss rates do not exceed $0.5\%$ of the AGN
luminosity. Thermal winds make only a small 
contribution to the AGN feedback and evolution of the host galaxy.
This is consistent with results of full radiation hydrodynamic models of X-ray heated 
thermal winds from the torus \citep{dor08a,dor08b}.


\section{Observational characteristics of thermal winds in AGN}

While the thermal wind does not carry enough energy to determine AGN feedback, it may still
produce the observed warm absorber features. \citet{don18} show that the 
column density and ionisation along the line of sight could be analytically estimated for
the black hole binary case, (i.e., when the wind is launched from the
disc and there are no obstacles such as torus in the line of sight). They could match to 
the results of radiative hydrodynamic models \citep{woo96} by assuming
radial streamlines and $n(R,\mu) =
n_0(R)(1-\mu)$, where $n_0$ is the number density for an edge-on inclination and $\mu=\cos i$.  
However, in the AGN case, the geometry is more complex,
with the BLR clouds and torus having some scale height so the wind 
does not extend down to edge-on inclination angles. Our  
torus geometry has opening angle of $56^\circ$, so 
we assume that the same amount of thermal wind gas exists within $0^\circ<i<56^\circ$, i.e.,
\begin{equation}
\int_0^1\!n_0(R)(1-\mu) d\mu 
= A \int_{\mu_0}^1\!n_0(R)(1-\mu) d\mu,
\end{equation}
where $\mu_0=\cos 56^\circ$. This gives a flat disc to funnel correction coefficient 
of $A=(1-\mu_0)^{-2}$.
The column density of the wind is then
\begin{equation}
\begin{split}
N_{\rm H}(\mu)&=A\int_0^{R_{\rm shield}}\! n_0(R)(1-\mu) dR\\
&=\frac{(1-\mu)}{(1-\mu_0)^2}\int_0^{R_{\rm shield}}\! \frac{\dot{m}(R)}{c_{\rm ch}(R)m_{\rm I}} dR, \label{eq:NH}
\end{split}
\end{equation}
where $m_{\rm I}=1.26m_{\rm p}$ is the mean ion mass per electron.

\begin{figure}
\centering
	\includegraphics[width=0.8\columnwidth, angle=270]{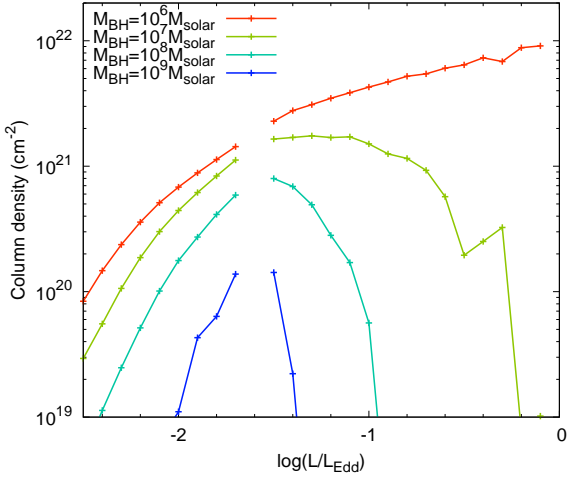}
    \caption{Column densities of the wind when $i=30^\circ$}
    \label{fig:column}
\end{figure}

\begin{figure}
\centering
	\includegraphics[width=0.8\columnwidth, angle=270]{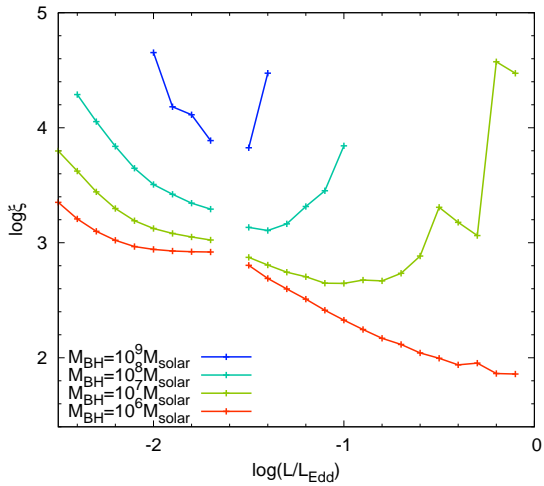}
    \caption{Ionisation parameters of the wind. The cases of $N_{\rm H}<10^{19}$~cm$^{-2}$ are not shown.}
    \label{fig:logxi}
\end{figure}

Fig.~\ref{fig:column} shows the column density predicted for
$i=30^\circ$ for a range of mass and mass accretion rates.  We
also calculate the ionisation parameter as $\xi(R)=L_{\rm
  ion}/n(R)R^2$, where $L_{\rm ion}$ is the luminosity from 13.6~eV to 13.6~keV. Our radial streamlines
  mean that this is constant with radius, with the value shown Fig.~\ref{fig:logxi}.  These
predictions result in typical columns of $10^{20-21}~\mathrm{cm}^{-2}$ at $\log\xi\sim 2-3$ for
black hole masses $<10^7\ M_\odot$, which are close to the observed
values for warm absorbers, as are the typical velocities of few hundred to a few thousand km s$^{-1}$
(see Fig.~\ref{fig:velocity}).

\section{Comparison to observations}

The predictions above for thermal winds appear to be quite well matched to the observed properties of warm absorbers. However, the individual AGNs all have different mass, luminosity, inclination, and
presumably spin.  This makes it difficult to compare the change in wind
properties using tracks of constant mass.  Therefore, we instead
make predictions for each individual system from its own estimated mass and luminosity. 
We use the radio quiet AGN sample of \citet{lah14}, which tabulates mass and luminosity along with 
the observed warm absorber properties, keeping our assumption of zero spin.

\begin{figure}
\centering
	\includegraphics[width=0.75\columnwidth, angle=270]{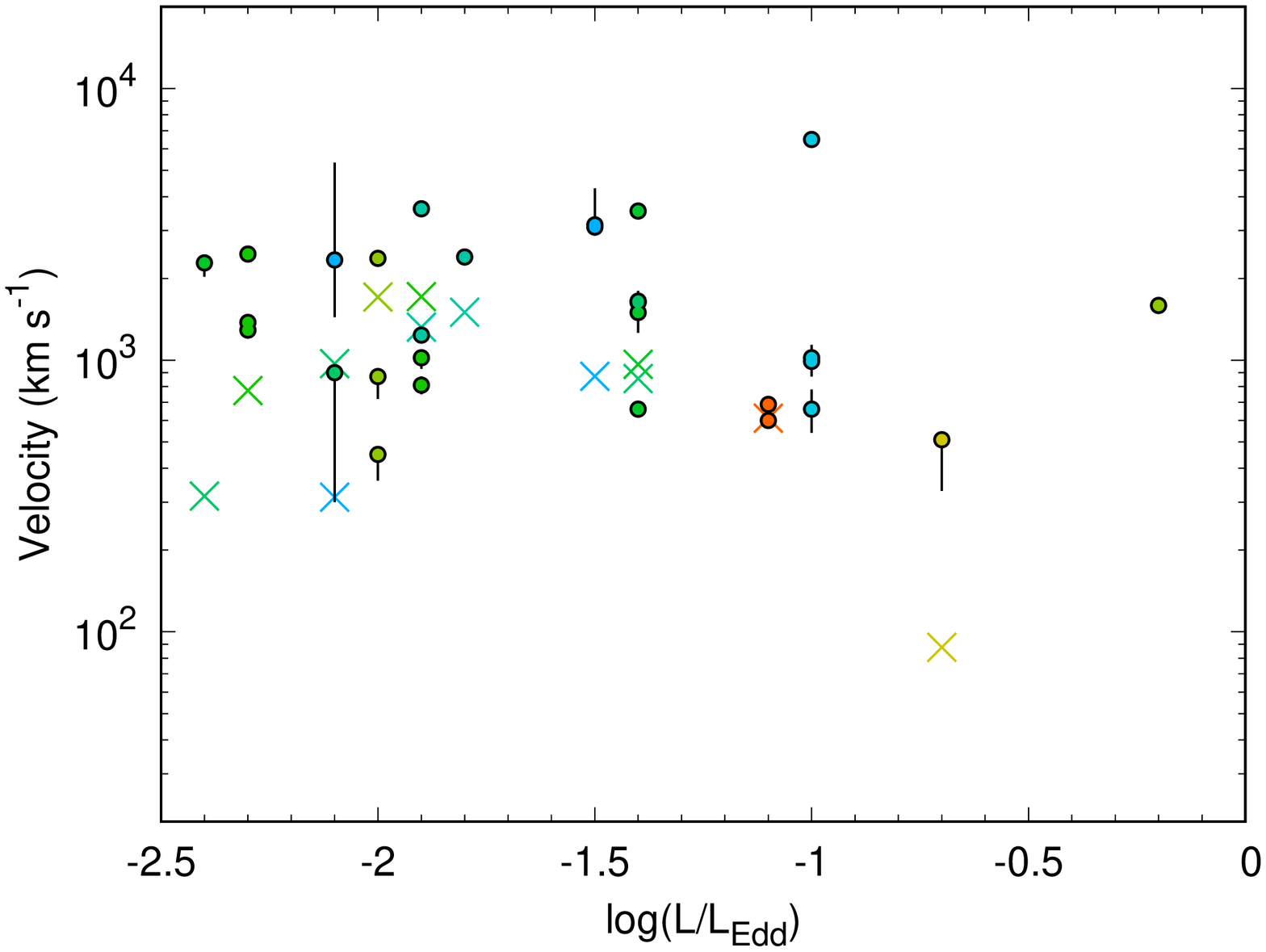}
	\includegraphics[width=0.75\columnwidth, angle=270]{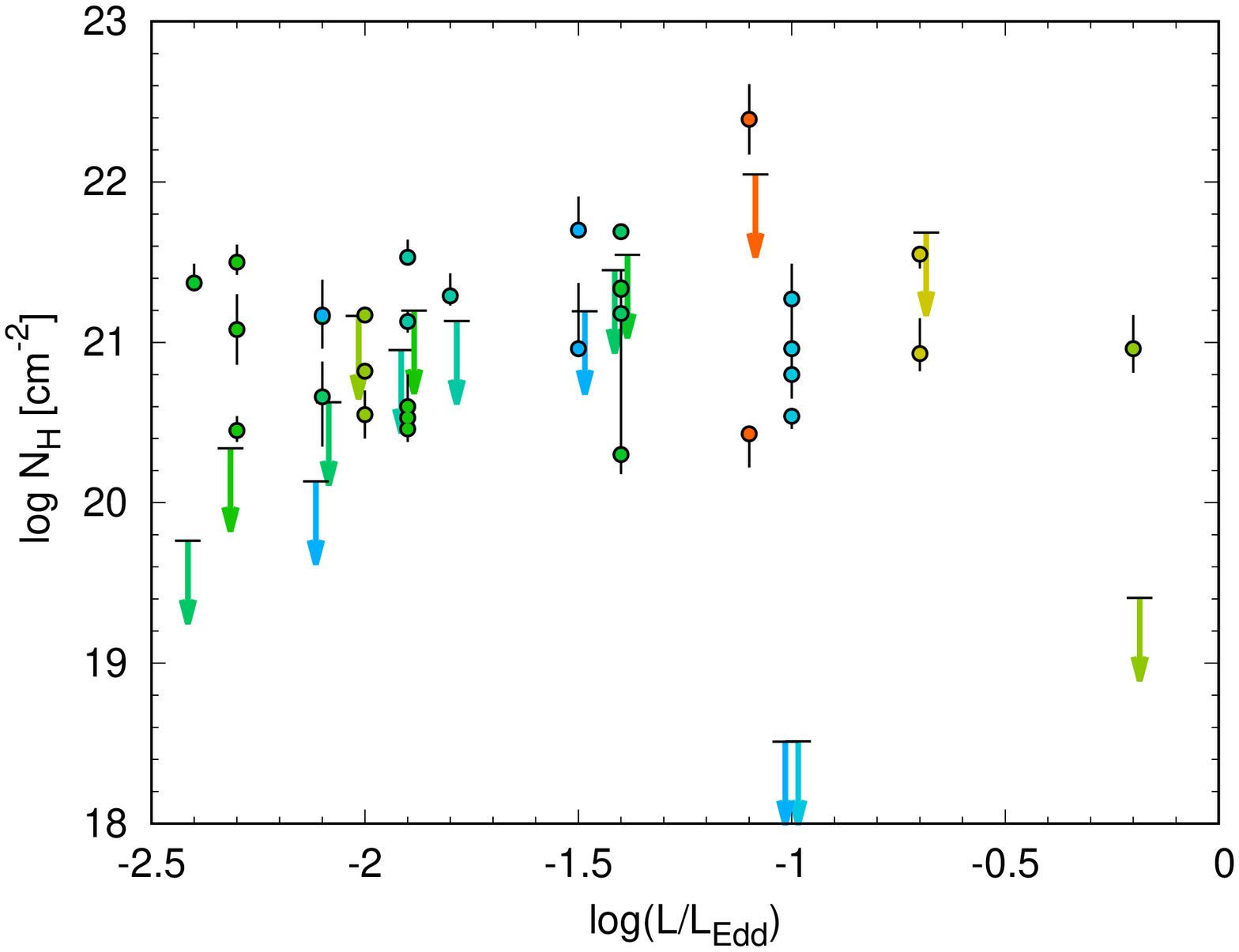}
	\includegraphics[width=0.75\columnwidth, angle=270]{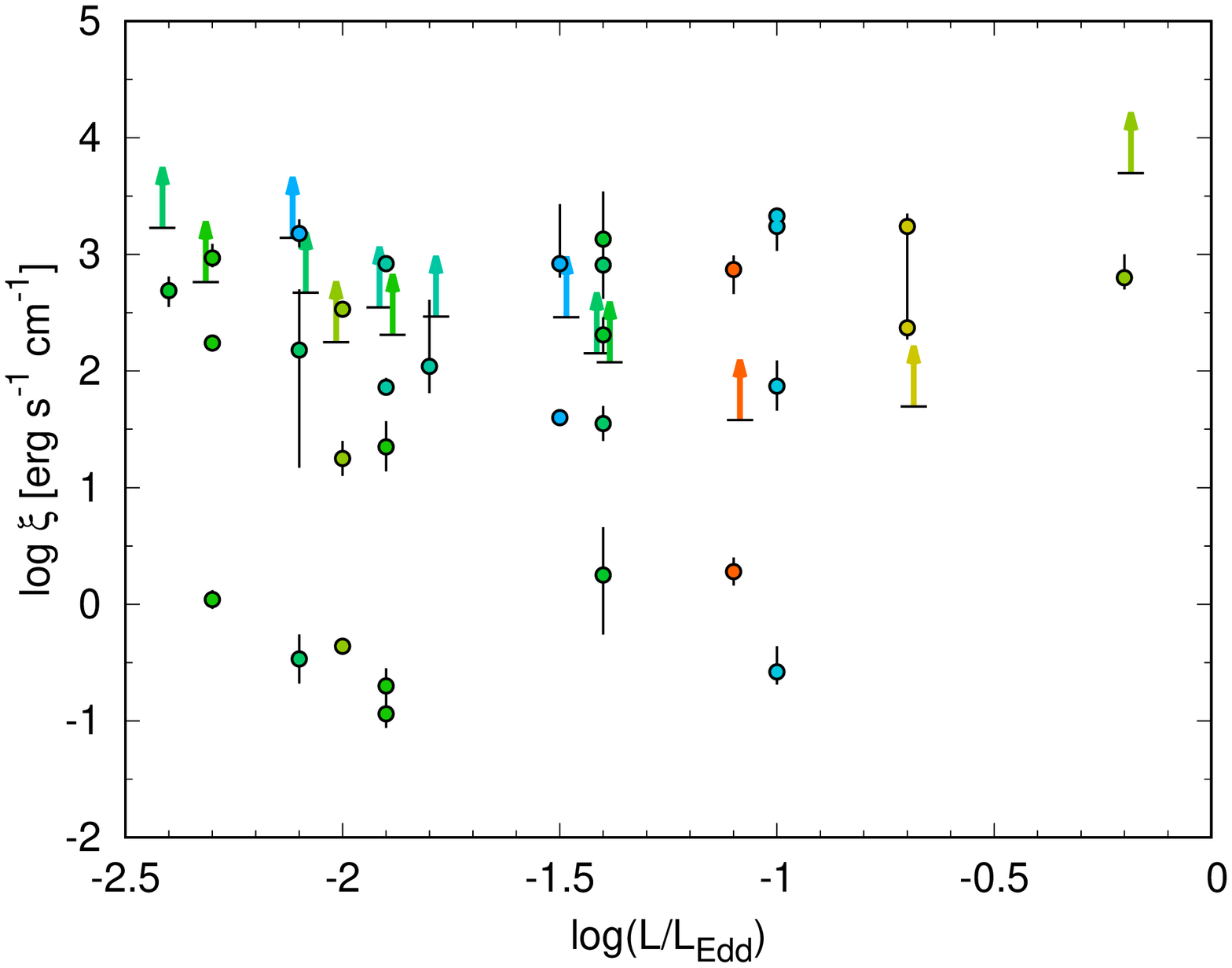}
    \caption{Eddington ratio dependence of warm absorber parameters shown in \citet{lah14}, with our models.
    Each model point corresponds to each black hole mass and Eddington ratio. The same colour show the same black hole mass, from red ($M_{\rm BH}=10^6\,M_\odot$) to blue ($M_{\rm BH}=10^9\,M_\odot$).
   Some targets have more observation points than the model ones because they have multiple warm absorber components and/or because the model points are outside of the shown region. In the middle and lower panels, the start points in the model arrows (with short black horizontal lines) show when $i=56^\circ$, and the end points show when $i=30^\circ$.
    }
    \label{fig:warmabs}
\end{figure}

The upper panel of Fig.~\ref{fig:warmabs} shows a comparison between
the observed wind velocities (circle bins with error bars) and the
velocities we have calculated (crosses).  The thermal wind model
velocities are clearly consistent with the observed velocities of
$v\sim1000$~km~s$^{-1}$.

Next, we compare the column density in the middle panel in
Fig.~\ref{fig:warmabs}.  The calculated $N_{\rm H}$ values depend on
the inclination angle from $0^\circ\leq i \leq 56^\circ$ (see equation \ref{eq:NH}), and thus we plot the upper limit (when $i=56^\circ$) as a vertical arrow.  
The end point of the arrow corresponds to $i=30^\circ$.
Again the agreement with the observed columns is generally good, with some exceptions.

Finally, we show our results for the ionisation parameter and the corresponding
observed values are shown in the bottom panel in
Fig.~\ref{fig:warmabs}, again with a vertical arrow marking the lower limit
of ionisation expected from the possible range of inclination angles.
Here the predicted values form an upper bound of the observed
ionisation states, but there are many points which lie 2--4
orders of magnitude below. Most of these are from the same objects as 
are also detected with absorption columns at the much higher ionisation stage, as 
most of the warm absorbers detected in AGN are multi-phase. 

\begin{figure}
\centering
	\includegraphics[width=0.8\columnwidth,angle=270]{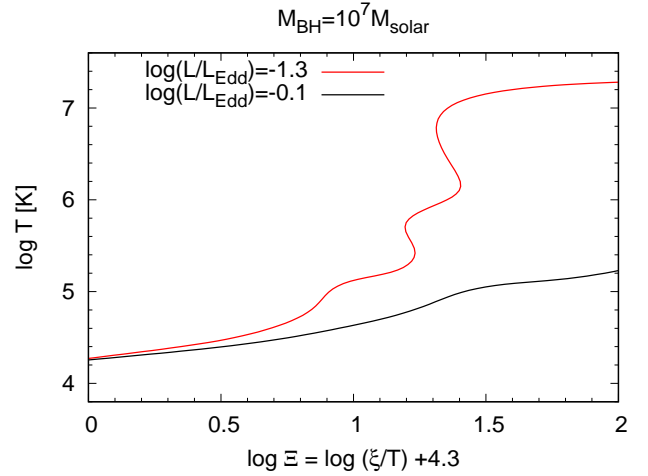}
    \caption{Thermal equilibrium curves for gas in pressure
balance for our SED model. The surface density and ionisation parameter are assumed to be $n=10^8$~cm$^{-3}$ and $\log\xi=5$, respectively.}
    \label{fig:Scurve}
\end{figure}

We first consider whether the thermal wind can naturally produce such a
multi-phase absorber. Studies of individual objects show that many of
these phases appear to be in pressure equilibrium (e.g.\ \citealt{kro03,kro07,kro09,net03}).  
The red line in Fig.~\ref{fig:Scurve}
shows the thermal equilibrium curves for gas in pressure
balance for our SED for $\log(L/L_{\rm Edd})=-1.3$ and
$M_{\rm BH}=10^7M_\odot$. Regions to the lower right of the line have
heating larger than cooling, while the upper left has cooling larger
than heating. Where the thermal stability curve has an S-bend, this
indicates that material on the middle branch is thermally unstable as
it will rapidly heat up to join the upper branch, or cool down to join
the lower branch.  Two phase gas can stably co-exist in pressure
equilibrium for $\Xi$ between the minimum value on the upper branch
and the maximum value on the lower \citep{kro81}. 
There are two regions of S curve for this SED, one at $\log\Xi\sim
1.3$ where the gas can be at $10^7$~K on the upper branch, or $10^6$~K
on the lower, and another separate region at $\log\Xi\sim 1.2$ where
the gas can be at $\sim 6\times 10^5$~K or $1.5\times 10^5$~K.  Since
pressure $\propto nT$ then these separate phases have densities which
are higher in proportion to their lower temperature, so they are
characterised by a standard ionisation parameter ($\xi=L/nR^2$) which
is lower \citep{roz06,gon06}.

\citet{hol07} introduced the absorption measure distribution (AMD) which
defines the distribution of column density as a function of the
standard ionisation parameter. They interpret a distinct dip in this
distribution for the warm absorber gas in several objects as evidence
for the thermal instability (see \citealt{beh09,hol12}).

\begin{figure}
\centering
	\includegraphics[width=0.8\columnwidth,angle=270]{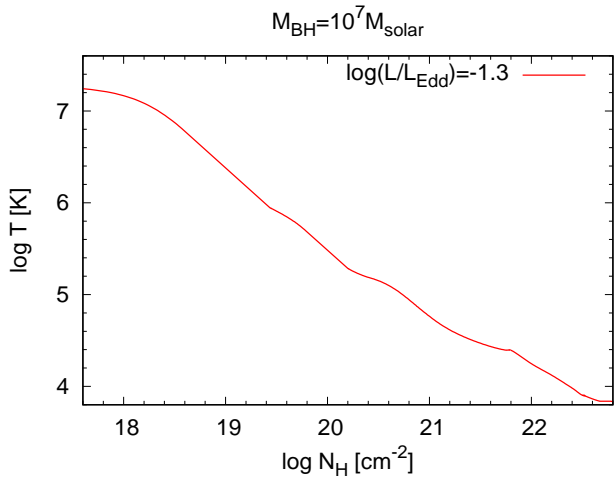}
    \caption{Temperature of the slab gas irradiated by our SED model for $M_{\rm BH}=10^7M_\odot$ and $\log(L/L_{\rm Edd})=-1.3$.
    The setting is same as Fig.~\ref{fig:Scurve}.}
    \label{fig:cooling}
\end{figure}

The thermal instabilities are always triggered in material with large
enough column density (radiation pressure confinement: 
\citealt{ste14,goo16,adh19}). Fig.\ref{fig:cooling} shows the
temperature versus column density into an irradiated slab in pressure balance for this  SED and black hole mass. Given that Fig.~\ref{fig:param}b shows that the luminosity is around the
critical luminosity, this means that the Compton heating time is only just short enough 
to get material onto the upper branch, i.e. that its maximum ionisation is quite close to the 
critical ionisation parameter $\Xi_{\rm max}\sim 40$. The material then 
starts out close to the instability point so a multi-phase outflow may quite easily
develop even with a total column of only $10^{21}$~cm$^{-2}$.

However, the more material drops down onto the lower branch, the less there is on the upper branch. The predicted column densities with all the material in the upper branch were already a little low, so 
putting more material on the lower branch to decrease the tension with the ionisation parameters will increase the tension with the column density. This is especially an issue with the high mass accretion rates, where the column density is predicted to be lower, and there is no true instability (black line in Fig.~\ref{fig:Scurve}). Instead, it seems more likely that there are additional processes at work
which enhance the column density of wind material.  

The thermal instability is also required to explain the UV absorption lines.
 In some AGNs (e.g.~NGC 4151) the lower ionisation phase of the warm absorber gas can be traced through both soft X-ray absorption and UV absorption. 
 The low ionisation parameter means that it cannot be on the upper branch of the thermal heating/cooling curve, so it has much lower temperature than required to escape as a thermal wind \citep{cre07}. This material could be accelerated via thermal driving to its current velocity, and then cool (and probably clump) via the thermal instability.

The requirement for additional wind material is seen even more clearly in
radio loud AGN. These can show warm absorbers, though the columns are generally towards the lower range of those seen in radio quiet AGN \citep{meh19}. However, the black holes are very massive ($M_{\rm BH}\sim10^{8-10}M_\odot$) where our model gives a thermal wind launching radius which exceeds the self-shielding radius of the torus (see Fig.~\ref{fig:param2}). The thermal wind column density is then negligible. 

\section{Discussion}\label{sec7}

Our thermal wind model successfully predicts the velocity of the warm absorbers (not the ultrafast outflows) seen in AGN, but the column density of material is often a little lower than observed especially if this material has to be shared between multiple phases of gas with different ionisation state. Instead, it seems more likely that there are additional sources of material and/or launch mechanisms for these outflows.
This is especially the case for higher mass AGN, where the launch radius for thermal winds is much larger than the inner edge of the dust torus. Our assumed circular torus geometry means that there is self-shielding at these radii, so very little wind is predicted. 
A different assumed torus geometry would change this, e.g. a 
torus with conical/flared rather than circular cross-section 
would allow material further out to be illuminated, increasing the total column density of the thermal wind predictions as these go as $N_{\rm H}\propto \log (R_{\rm out}/R_{\rm in})$. Similarly, a clumpy rather than smooth torus (e.g.~\citealt{nen08}) will allow more irradiation at larger radii as well as make a more inhomogeneous thermal wind. 

The torus shape should be determined self consistently, as it is sculpted by the 
mass loss via the wind. Hydrodynamic simulations of this are shown in \citet{dor08a,dor08b} and \citep{dor09}, where they start with a smooth torus of finite optical depth, and follow the evolution as the X-ray heating produces a thermal wind. However, in these simulations, the inner edge of the torus is arbitrary rather than being tied to the dust sublimation radius. They assume 
a torus central radius of  $R_0=0.5-1.5$~pc, with shape such that the inner edge is at $0.23R_0$ (for their $d=2.5$), which is 
much larger than the inner edge of the dust sublimation region at 0.02--0.04~pc for
their black hole mass of $10^6M_\odot$ and assumed $L/L_{\rm Edd}=0.1-0.5$ (see Fig.~\ref{fig:param2}). 
They predict large column densities of gas at high inclinations, but much of this material is part of the torus rather than being in a wind as their torus density distribution means that it is optically thin around its edge. This is more realistic than the infinitely optically thick torus used here, but cannot explain the observed {\it outflow} columns unless the torus (and/or BLR) is itself part of an additional wind rather than just rotating as assumed here. Nor can it explain the observed statistics of the incidence of warm absorbers in more than 50\% of AGN as these large columns only cover a small fraction of the torus opening \citep{dor09}. The question then becomes what drives the wind which provides the material for the torus and BLR. 

Dust is an obvious component of the torus, and may well also be responsible for the underlying driving mechanism 
for the BLR \citep{cze11}.
Dust opacity is much larger than the electron scattering cross-section,
particularly for UV wavelengths, 
so there is again a `force multiplier' effect, reducing the effective Eddington limit. 
\citet{fab08} showed that this is $\ge 100$ for 
realistic AGN SEDs for columns $\lesssim 10^{21}$~cm$^{-2}$.
Thus all of the bright AGN considered here are effectively super-Eddington for low columns of dusty gas, so should power a wind from the surface of the torus. Hydrodynamic simulations including dust have focused on the role of trapped infrared radiation in producing the scale height of the torus \citep{dor12a,dor12b}, but models including  dynamics show a strong dust driven wind from the inner edge of the torus \citep{wad12,wad15,cha16,dor16}.

Hence it seems most likely that the warm absorbers are thermal-radiative winds, rather than solely thermal, and where the radiation pressure is mainly on the dust grains present in the torus. 

\section{Conclusion}

\citet{kro01} showed that the warm absorbers seen in AGN
result could plausibly result from a thermal wind from the torus, and
highlighted the role of the thermal instability in producing the
observed multi-phase absorbing gas. Thermal winds are sensitive to the
SED, and there is now more observational data on how the SED 
changes as a
function of mass and mass accretion rate in AGN. We use the specific
models of \citet{kub18} which can reproduce these general
trends ({\sc qsosed}). These predict that the Compton temperature
decreases with Eddington ratio, which increases the radius from which
the wind is launched in terms of $R/R_g\propto \dot{m}^{-1}$. The irradiated material in the BLR and torus has radius $R/R_g\propto (\dot{m}/M)^{1/2}$ so thermal winds are launched from the BLR for low mass, low mass accretion rate AGN, and from the outer edge of the torus for high mass, high mass accretion rates. We use the analytic models of thermal winds developed by \citet{don18} to make quantitative predictions of the amount of column produced by thermal winds in AGN.
We show generally that these predictions are in tension with the observations, especially if the thermal instability is invoked in order to produce the observed low ionisation gas components. Instead, dust driven winds from the inner edge of the torus is more likely to be the origin of much of the observed outflow material, with the X-ray heating contributing more to the observed multi-phase ionisation structure. 
Hydrodynamic simulations of this complex, dynamic environment
show that this produces outflows and obscuration from the torus
in line with the observations \citep{wad15,cha16,dor16}. 

Magnetic winds are not required in this picture. There are two main arguments used as evidence for such winds. 
In NGC 4151 (and other well studies AGN) the lower ionisation phase traced by UV absorption must have much lower temperature than required to escape as a thermal wind \citep{cre07}. This is true, but this material could be accelerated via either thermal or dust driving and then cool (and probably clump) via the thermal instability.
The second piece of evidence is that there is an
anti-correlation of warm absorber column density with radio loudness in a sample of radio loud AGN \citep{meh19}. This could link the winds causally to the jet, which is clearly a magnetic structure and similar anti-correlation of wind and jet is seen directly in the stellar mass black hole binaries \citep{nei09,pon12}. Nonetheless, in binaries the link to the jet is more likely to be via the changing SED with changing spectral state rather than causally via the jet magnetic field \citep{tom19}. The origin of the correlation in the small sample of radio loud AGN is yet to be clarified, but dust and thermal driving together clearly have the potential to explain most of the warm absorbers seen.

\section*{Acknowledgements}
MM is financially supported by Japan Society for the Promotion of Science (JSPS) Overseas Research Fellowship.
This work is partly supported by the Science and Technology Facilities Council (STFC) grant ST/P000541/1 (CD), 
Kavli Institute for the Physics and Mathematics of the Universe (IPMU) funding from the National Science Foundation No.\ NSF PHY17-48958 (CD), and JSPS KAKENHI Grant Number 19J13373 (RT).



\bibliographystyle{mnras}
\bibliography{00} 

\begin{thebibliography}{}
\makeatletter
\relax
\def\mn@urlcharsother{\let\do\@makeother \do\$\do\&\do\#\do\^\do\_\do\%\do\~}
\def\mn@doi{\begingroup\mn@urlcharsother \@ifnextchar [ {\mn@doi@}
  {\mn@doi@[]}}
\def\mn@doi@[#1]#2{\def\@tempa{#1}\ifx\@tempa\@empty \href
  {http://dx.doi.org/#2} {doi:#2}\else \href {http://dx.doi.org/#2} {#1}\fi
  \endgroup}
\def\mn@eprint#1#2{\mn@eprint@#1:#2::\@nil}
\def\mn@eprint@arXiv#1{\href {http://arxiv.org/abs/#1} {{\tt arXiv:#1}}}
\def\mn@eprint@dblp#1{\href {http://dblp.uni-trier.de/rec/bibtex/#1.xml}
  {dblp:#1}}
\def\mn@eprint@#1:#2:#3:#4\@nil{\def\@tempa {#1}\def\@tempb {#2}\def\@tempc
  {#3}\ifx \@tempc \@empty \let \@tempc \@tempb \let \@tempb \@tempa \fi \ifx
  \@tempb \@empty \def\@tempb {arXiv}\fi \@ifundefined
  {mn@eprint@\@tempb}{\@tempb:\@tempc}{\expandafter \expandafter \csname
  mn@eprint@\@tempb\endcsname \expandafter{\@tempc}}}

\bibitem[\protect\citeauthoryear{{Adhikari}, {R{\'o}{\.z}a{\'n}ska},
  {Hryniewicz}, {Czerny}  \& {Behar}}{{Adhikari} et~al.}{2019}]{adh19}
{Adhikari} T.~P.,  {R{\'o}{\.z}a{\'n}ska} A.,  {Hryniewicz} K.,  {Czerny} B.,
  {Behar} E.,  2019, \apj, \href
  {https://ui.adsabs.harvard.edu/abs/2018arXiv181205154A} {submitted (arXiv:
  1812.05154)}

\bibitem[\protect\citeauthoryear{{Balsara} \& {Krolik}}{{Balsara} \&
  {Krolik}}{1993}]{bal93}
{Balsara} D.~S.,  {Krolik} J.~H.,  1993, \mn@doi [\apj] {10.1086/172116}, \href
  {https://ui.adsabs.harvard.edu/abs/1993ApJ...402..109B} {402, 109}

\bibitem[\protect\citeauthoryear{{Barvainis}}{{Barvainis}}{1987}]{bar87}
{Barvainis} R.,  1987, \mn@doi [\apj] {10.1086/165571}, \href
  {http://adsabs.harvard.edu/abs/1987ApJ...320..537B} {320, 537}

\bibitem[\protect\citeauthoryear{{Begelman}, {McKee}  \& {Shields}}{{Begelman}
  et~al.}{1983}]{beg83}
{Begelman} M.~C.,  {McKee} C.~F.,   {Shields} G.~A.,  1983, \mn@doi [\apj]
  {10.1086/161178}, \href {http://adsabs.harvard.edu/abs/1983ApJ...271...70B}
  {271, 70}

\bibitem[\protect\citeauthoryear{{Behar}}{{Behar}}{2009}]{beh09}
{Behar} E.,  2009, \mn@doi [\apj] {10.1088/0004-637X/703/2/1346}, \href
  {https://ui.adsabs.harvard.edu/abs/2009ApJ...703.1346B} {703, 1346}

\bibitem[\protect\citeauthoryear{{Bentz} et~al.,}{{Bentz} et~al.}{2013}]{ben13}
{Bentz} M.~C.,  et~al., 2013, \mn@doi [\apj] {10.1088/0004-637X/767/2/149},
  \href {http://adsabs.harvard.edu/abs/2013ApJ...767..149B} {767, 149}

\bibitem[\protect\citeauthoryear{{Blandford} \& {Payne}}{{Blandford} \&
  {Payne}}{1982}]{bla82}
{Blandford} R.~D.,  {Payne} D.~G.,  1982, \mn@doi [\mnras]
  {10.1093/mnras/199.4.883}, \href
  {http://adsabs.harvard.edu/abs/1982MNRAS.199..883B} {199, 883}

\bibitem[\protect\citeauthoryear{{Blustin}, {Page}, {Fuerst}, {Branduardi-
  Raymont}  \& {Ashton}}{{Blustin} et~al.}{2005}]{blu05}
{Blustin} A.~J.,  {Page} M.~J.,  {Fuerst} S.~V.,  {Branduardi- Raymont} G.,
  {Ashton} C.~E.,  2005, \mn@doi [\aap] {10.1051/0004-6361:20041775}, \href
  {https://ui.adsabs.harvard.edu/#abs/2005A&A...431..111B} {431, 111}

\bibitem[\protect\citeauthoryear{{Chan} \& {Krolik}}{{Chan} \&
  {Krolik}}{2016}]{cha16}
{Chan} C.-H.,  {Krolik} J.~H.,  2016, \mn@doi [\apj]
  {10.3847/0004-637X/825/1/67}, \href
  {https://ui.adsabs.harvard.edu/abs/2016ApJ...825...67C} {825, 67}

\bibitem[\protect\citeauthoryear{{Crenshaw} \& {Kraemer}}{{Crenshaw} \&
  {Kraemer}}{2007}]{cre07}
{Crenshaw} D.~M.,  {Kraemer} S.~B.,  2007, \mn@doi [\apj] {10.1086/511970},
  \href {https://ui.adsabs.harvard.edu/abs/2007ApJ...659..250C} {659, 250}

\bibitem[\protect\citeauthoryear{{Czerny} \& {Hryniewicz}}{{Czerny} \&
  {Hryniewicz}}{2011}]{cze11}
{Czerny} B.,  {Hryniewicz} K.,  2011, \mn@doi [\aap]
  {10.1051/0004-6361/201016025}, \href
  {https://ui.adsabs.harvard.edu/abs/2011A&A...525L...8C} {525, L8}

\bibitem[\protect\citeauthoryear{{Done}, {Davis}, {Jin}, {Blaes}  \&
  {Ward}}{{Done} et~al.}{2012}]{don12}
{Done} C.,  {Davis} S.~W.,  {Jin} C.,  {Blaes} O.,   {Ward} M.,  2012, \mn@doi
  [\mnras] {10.1111/j.1365-2966.2011.19779.x}, \href
  {https://ui.adsabs.harvard.edu/abs/2012MNRAS.420.1848D} {420, 1848}

\bibitem[\protect\citeauthoryear{{Done}, {Tomaru}  \& {Takahashi}}{{Done}
  et~al.}{2018}]{don18}
{Done} C.,  {Tomaru} R.,   {Takahashi} T.,  2018, \mn@doi [\mnras]
  {10.1093/mnras/stx2400}, \href
  {http://adsabs.harvard.edu/abs/2018MNRAS.473..838D} {473, 838}

\bibitem[\protect\citeauthoryear{{Dorodnitsyn} \& {Kallman}}{{Dorodnitsyn} \&
  {Kallman}}{2009}]{dor09}
{Dorodnitsyn} A.,  {Kallman} T.,  2009, \mn@doi [\apj]
  {10.1088/0004-637X/703/2/1797}, \href
  {https://ui.adsabs.harvard.edu/abs/2009ApJ...703.1797D} {703, 1797}

\bibitem[\protect\citeauthoryear{{Dorodnitsyn} \& {Kallman}}{{Dorodnitsyn} \&
  {Kallman}}{2012}]{dor12b}
{Dorodnitsyn} A.,  {Kallman} T.,  2012, \mn@doi [\apj]
  {10.1088/0004-637X/761/1/70}, \href
  {https://ui.adsabs.harvard.edu/abs/2012ApJ...761...70D} {761, 70}

\bibitem[\protect\citeauthoryear{{Dorodnitsyn}, {Kallman}  \&
  {Proga}}{{Dorodnitsyn} et~al.}{2008a}]{dor08a}
{Dorodnitsyn} A.,  {Kallman} T.,   {Proga} D.,  2008a, \mn@doi [\apjl]
  {10.1086/529374}, \href {http://adsabs.harvard.edu/abs/2008ApJ...675L...5D}
  {675, L5}

\bibitem[\protect\citeauthoryear{{Dorodnitsyn}, {Kallman}  \&
  {Proga}}{{Dorodnitsyn} et~al.}{2008b}]{dor08b}
{Dorodnitsyn} A.,  {Kallman} T.,   {Proga} D.,  2008b, \mn@doi [\apj]
  {10.1086/591418}, \href
  {https://ui.adsabs.harvard.edu/abs/2008ApJ...687...97D} {687, 97}

\bibitem[\protect\citeauthoryear{{Dorodnitsyn}, {Kallman}  \&
  {Bisnovatyi-Kogan}}{{Dorodnitsyn} et~al.}{2012}]{dor12a}
{Dorodnitsyn} A.,  {Kallman} T.,   {Bisnovatyi-Kogan} G.~S.,  2012, \mn@doi
  [\apj] {10.1088/0004-637X/747/1/8}, \href
  {https://ui.adsabs.harvard.edu/abs/2012ApJ...747....8D} {747, 8}

\bibitem[\protect\citeauthoryear{{Dorodnitsyn}, {Kallman}  \&
  {Proga}}{{Dorodnitsyn} et~al.}{2016}]{dor16}
{Dorodnitsyn} A.,  {Kallman} T.,   {Proga} D.,  2016, \mn@doi [\apj]
  {10.3847/0004-637X/819/2/115}, \href
  {https://ui.adsabs.harvard.edu/abs/2016ApJ...819..115D} {819, 115}

\bibitem[\protect\citeauthoryear{{Elitzur} \& {Shlosman}}{{Elitzur} \&
  {Shlosman}}{2006}]{eli06}
{Elitzur} M.,  {Shlosman} I.,  2006, \mn@doi [\apjl] {10.1086/508158}, \href
  {https://ui.adsabs.harvard.edu/abs/2006ApJ...648L.101E} {648, L101}

\bibitem[\protect\citeauthoryear{{Fabian}, {Vasudevan}  \& {Gandhi}}{{Fabian}
  et~al.}{2008}]{fab08}
{Fabian} A.~C.,  {Vasudevan} R.~V.,   {Gandhi} P.,  2008, \mn@doi [\mnras]
  {10.1111/j.1745-3933.2008.00430.x}, \href
  {https://ui.adsabs.harvard.edu/abs/2008MNRAS.385L..43F} {385, L43}

\bibitem[\protect\citeauthoryear{{Fukumura}, {Kazanas}, {Contopoulos}  \&
  {Behar}}{{Fukumura} et~al.}{2010}]{fuk10}
{Fukumura} K.,  {Kazanas} D.,  {Contopoulos} I.,   {Behar} E.,  2010, \mn@doi
  [\apj] {10.1088/0004-637X/715/1/636}, \href
  {http://adsabs.harvard.edu/abs/2010ApJ...715..636F} {715, 636}

\bibitem[\protect\citeauthoryear{{Gon{\c{c}}alves}, {Collin}, {Dumont},
  {Mouchet}, {R{\'o}{\.z}a{\'n}ska}, {Chevallier}  \&
  {Goosmann}}{{Gon{\c{c}}alves} et~al.}{2006}]{gon06}
{Gon{\c{c}}alves} A.~C.,  {Collin} S.,  {Dumont} A.~M.,  {Mouchet} M.,
  {R{\'o}{\.z}a{\'n}ska} A.,  {Chevallier} L.,   {Goosmann} R.~W.,  2006,
  \mn@doi [\aap] {10.1051/0004-6361:20064849}, \href
  {https://ui.adsabs.harvard.edu/abs/2006A&A...451L..23G} {451, L23}

\bibitem[\protect\citeauthoryear{{Goosmann}, {Holczer}, {Mouchet}, {Dumont},
  {Behar}, {Godet}, {Gon{\c{c}}alves}  \& {Kaspi}}{{Goosmann}
  et~al.}{2016}]{goo16}
{Goosmann} R.~W.,  {Holczer} T.,  {Mouchet} M.,  {Dumont} A.~M.,  {Behar} E.,
  {Godet} O.,  {Gon{\c{c}}alves} A.~C.,   {Kaspi} S.,  2016, \mn@doi [\aap]
  {10.1051/0004-6361/201425199}, \href
  {https://ui.adsabs.harvard.edu/abs/2016A&A...589A..76G} {589, A76}

\bibitem[\protect\citeauthoryear{{Higginbottom}, {Knigge}, {Long}, {Matthews},
  {Sim}  \& {Hewitt}}{{Higginbottom} et~al.}{2018}]{hig18}
{Higginbottom} N.,  {Knigge} C.,  {Long} K.~S.,  {Matthews} J.~H.,  {Sim}
  S.~A.,   {Hewitt} H.~A.,  2018, \mn@doi [\mnras] {10.1093/mnras/sty1599},
  \href {https://ui.adsabs.harvard.edu/abs/2018MNRAS.479.3651H} {479, 3651}

\bibitem[\protect\citeauthoryear{{Holczer} \& {Behar}}{{Holczer} \&
  {Behar}}{2012}]{hol12}
{Holczer} T.,  {Behar} E.,  2012, \mn@doi [\apj] {10.1088/0004-637X/747/1/71},
  \href {https://ui.adsabs.harvard.edu/abs/2012ApJ...747...71H} {747, 71}

\bibitem[\protect\citeauthoryear{{Holczer}, {Behar}  \& {Kaspi}}{{Holczer}
  et~al.}{2007}]{hol07}
{Holczer} T.,  {Behar} E.,   {Kaspi} S.,  2007, \mn@doi [\apj]
  {10.1086/518416}, \href
  {https://ui.adsabs.harvard.edu/abs/2007ApJ...663..799H} {663, 799}

\bibitem[\protect\citeauthoryear{{Hopkins} \& {Elvis}}{{Hopkins} \&
  {Elvis}}{2010}]{hop10}
{Hopkins} P.~F.,  {Elvis} M.,  2010, \mn@doi [\mnras]
  {10.1111/j.1365-2966.2009.15643.x}, \href
  {https://ui.adsabs.harvard.edu/abs/2010MNRAS.401....7H} {401, 7}

\bibitem[\protect\citeauthoryear{{Jin}, {Ward}, {Done}  \& {Gelbord}}{{Jin}
  et~al.}{2012}]{jin12}
{Jin} C.,  {Ward} M.,  {Done} C.,   {Gelbord} J.,  2012, \mn@doi [\mnras]
  {10.1111/j.1365-2966.2011.19805.x}, \href
  {https://ui.adsabs.harvard.edu/abs/2012MNRAS.420.1825J} {420, 1825}

\bibitem[\protect\citeauthoryear{{Kaastra}, {Mewe}, {Liedahl}, {Komossa}  \&
  {Brinkman}}{{Kaastra} et~al.}{2000}]{kaa00}
{Kaastra} J.~S.,  {Mewe} R.,  {Liedahl} D.~A.,  {Komossa} S.,   {Brinkman}
  A.~C.,  2000, \aap, \href
  {https://ui.adsabs.harvard.edu/abs/2000A&A...354L..83K} {354, L83}

\bibitem[\protect\citeauthoryear{{Kaspi}, {Brandt}, {Netzer}, {Sambruna},
  {Chartas}, {Garmire}  \& {Nousek}}{{Kaspi} et~al.}{2000}]{kas00b}
{Kaspi} S.,  {Brandt} W.~N.,  {Netzer} H.,  {Sambruna} R.,  {Chartas} G.,
  {Garmire} G.~P.,   {Nousek} J.~A.,  2000, \mn@doi [\apj] {10.1086/312697},
  \href {https://ui.adsabs.harvard.edu/abs/2000ApJ...535L..17K} {535, L17}

\bibitem[\protect\citeauthoryear{{Konigl} \& {Kartje}}{{Konigl} \&
  {Kartje}}{1994}]{kon94}
{Konigl} A.,  {Kartje} J.~F.,  1994, \mn@doi [\apj] {10.1086/174746}, \href
  {https://ui.adsabs.harvard.edu/abs/1994ApJ...434..446K} {434, 446}

\bibitem[\protect\citeauthoryear{{Krolik} \& {Kriss}}{{Krolik} \&
  {Kriss}}{1995}]{kro95}
{Krolik} J.~H.,  {Kriss} G.~A.,  1995, \mn@doi [\apj] {10.1086/175896}, \href
  {https://ui.adsabs.harvard.edu/abs/1995ApJ...447..512K} {447, 512}

\bibitem[\protect\citeauthoryear{{Krolik} \& {Kriss}}{{Krolik} \&
  {Kriss}}{2001}]{kro01}
{Krolik} J.~H.,  {Kriss} G.~A.,  2001, \mn@doi [\apj] {10.1086/323442}, \href
  {https://ui.adsabs.harvard.edu/abs/2001ApJ...561..684K} {561, 684}

\bibitem[\protect\citeauthoryear{{Krolik}, {McKee}  \& {Tarter}}{{Krolik}
  et~al.}{1981}]{kro81}
{Krolik} J.~H.,  {McKee} C.~F.,   {Tarter} C.~B.,  1981, \mn@doi [\apj]
  {10.1086/159303}, \href
  {https://ui.adsabs.harvard.edu/abs/1981ApJ...249..422K} {249, 422}

\bibitem[\protect\citeauthoryear{{Krongold}, {Nicastro}, {Brickhouse}, {Elvis},
  {Liedahl}  \& {Mathur}}{{Krongold} et~al.}{2003}]{kro03}
{Krongold} Y.,  {Nicastro} F.,  {Brickhouse} N.~S.,  {Elvis} M.,  {Liedahl}
  D.~A.,   {Mathur} S.,  2003, \mn@doi [\apj] {10.1086/378639}, \href
  {https://ui.adsabs.harvard.edu/abs/2003ApJ...597..832K} {597, 832}

\bibitem[\protect\citeauthoryear{{Krongold}, {Nicastro}, {Elvis}, {Brickhouse},
  {Binette}, {Mathur}  \& {Jim{\'e}nez-Bail{\'o}n}}{{Krongold}
  et~al.}{2007}]{kro07}
{Krongold} Y.,  {Nicastro} F.,  {Elvis} M.,  {Brickhouse} N.,  {Binette} L.,
  {Mathur} S.,   {Jim{\'e}nez-Bail{\'o}n} E.,  2007, \mn@doi [\apj]
  {10.1086/512476}, \href
  {https://ui.adsabs.harvard.edu/abs/2007ApJ...659.1022K} {659, 1022}

\bibitem[\protect\citeauthoryear{{Krongold} et~al.,}{{Krongold}
  et~al.}{2009}]{kro09}
{Krongold} Y.,  et~al., 2009, \mn@doi [\apj] {10.1088/0004-637X/690/1/773},
  \href {https://ui.adsabs.harvard.edu/abs/2009ApJ...690..773K} {690, 773}

\bibitem[\protect\citeauthoryear{{Kubota} \& {Done}}{{Kubota} \&
  {Done}}{2018}]{kub18}
{Kubota} A.,  {Done} C.,  2018, \mn@doi [\mnras] {10.1093/mnras/sty1890}, \href
  {https://ui.adsabs.harvard.edu/#abs/2018MNRAS.480.1247K} {480, 1247}

\bibitem[\protect\citeauthoryear{{Laha}, {Guainazzi}, {Dewangan}, {Chakravorty}
   \& {Kembhavi}}{{Laha} et~al.}{2014}]{lah14}
{Laha} S.,  {Guainazzi} M.,  {Dewangan} G.~C.,  {Chakravorty} S.,   {Kembhavi}
  A.~K.,  2014, \mn@doi [\mnras] {10.1093/mnras/stu669}, \href
  {http://adsabs.harvard.edu/abs/2014MNRAS.441.2613L} {441, 2613}

\bibitem[\protect\citeauthoryear{{Laor} \& {Netzer}}{{Laor} \&
  {Netzer}}{1989}]{lao89}
{Laor} A.,  {Netzer} H.,  1989, \mn@doi [\mnras] {10.1093/mnras/238.3.897},
  \href {http://adsabs.harvard.edu/abs/1989MNRAS.238..897L} {238, 897}

\bibitem[\protect\citeauthoryear{{McKernan}, {Yaqoob}  \&
  {Reynolds}}{{McKernan} et~al.}{2007}]{mck07}
{McKernan} B.,  {Yaqoob} T.,   {Reynolds} C.~S.,  2007, \mn@doi [\mnras]
  {10.1111/j.1365-2966.2007.11993.x}, \href
  {http://adsabs.harvard.edu/abs/2007MNRAS.379.1359M} {379, 1359}

\bibitem[\protect\citeauthoryear{{Mehdipour} \& {Costantini}}{{Mehdipour} \&
  {Costantini}}{2019}]{meh19}
{Mehdipour} M.,  {Costantini} E.,  2019, \mn@doi [\aap]
  {10.1051/0004-6361/201935205}, \href
  {https://ui.adsabs.harvard.edu/abs/2019A&A...625A..25M} {625, A25}

\bibitem[\protect\citeauthoryear{{Neilsen} \& {Lee}}{{Neilsen} \&
  {Lee}}{2009}]{nei09}
{Neilsen} J.,  {Lee} J.~C.,  2009, \mn@doi [\nat] {10.1038/nature07680}, \href
  {https://ui.adsabs.harvard.edu/abs/2009Natur.458..481N} {458, 481}

\bibitem[\protect\citeauthoryear{{Nenkova}, {Sirocky}, {Ivezi{\'c}}  \&
  {Elitzur}}{{Nenkova} et~al.}{2008}]{nen08}
{Nenkova} M.,  {Sirocky} M.~M.,  {Ivezi{\'c}} {\v{Z}}.,   {Elitzur} M.,  2008,
  \mn@doi [\apj] {10.1086/590482}, \href
  {https://ui.adsabs.harvard.edu/abs/2008ApJ...685..147N} {685, 147}

\bibitem[\protect\citeauthoryear{{Netzer} et~al.,}{{Netzer}
  et~al.}{2003}]{net03}
{Netzer} H.,  et~al., 2003, \mn@doi [\apj] {10.1086/379508}, \href
  {https://ui.adsabs.harvard.edu/abs/2003ApJ...599..933N} {599, 933}

\bibitem[\protect\citeauthoryear{{Ponti}, {Fender}, {Begelman}, {Dunn},
  {Neilsen}  \& {Coriat}}{{Ponti} et~al.}{2012}]{pon12}
{Ponti} G.,  {Fender} R.~P.,  {Begelman} M.~C.,  {Dunn} R.~J.~H.,  {Neilsen}
  J.,   {Coriat} M.,  2012, \mn@doi [\mnras]
  {10.1111/j.1745-3933.2012.01224.x}, \href
  {https://ui.adsabs.harvard.edu/abs/2012MNRAS.422L..11P} {422, L11}

\bibitem[\protect\citeauthoryear{{Proga} \& {Kallman}}{{Proga} \&
  {Kallman}}{2004}]{pro04}
{Proga} D.,  {Kallman} T.~R.,  2004, \mn@doi [\apj] {10.1086/425117}, \href
  {http://adsabs.harvard.edu/abs/2004ApJ...616..688P} {616, 688}

\bibitem[\protect\citeauthoryear{{R{\'o}{\.z}a{\'n}ska}, {Goosmann}, {Dumont}
  \& {Czerny}}{{R{\'o}{\.z}a{\'n}ska} et~al.}{2006}]{roz06}
{R{\'o}{\.z}a{\'n}ska} A.,  {Goosmann} R.,  {Dumont} A.~M.,   {Czerny} B.,
  2006, \mn@doi [\aap] {10.1051/0004-6361:20052723}, \href
  {https://ui.adsabs.harvard.edu/abs/2006A&A...452....1R} {452, 1}

\bibitem[\protect\citeauthoryear{{Sako} et~al.,}{{Sako} et~al.}{2001}]{sak01}
{Sako} M.,  et~al., 2001, \mn@doi [\aap] {10.1051/0004-6361:20000081}, \href
  {https://ui.adsabs.harvard.edu/abs/2001A&A...365L.168S} {365, L168}

\bibitem[\protect\citeauthoryear{{Shakura} \& {Sunyaev}}{{Shakura} \&
  {Sunyaev}}{1973}]{sha73}
{Shakura} N.~I.,  {Sunyaev} R.~A.,  1973, \aap, \href
  {http://adsabs.harvard.edu/abs/1973A%26A....24..337S} {24, 337}

\bibitem[\protect\citeauthoryear{{Starling}, {Siemiginowska}, {Uttley}  \&
  {Soria}}{{Starling} et~al.}{2004}]{sta04}
{Starling} R.~L.~C.,  {Siemiginowska} A.,  {Uttley} P.,   {Soria} R.,  2004,
  \mn@doi [\mnras] {10.1111/j.1365-2966.2004.07167.x}, \href
  {http://adsabs.harvard.edu/abs/2004MNRAS.347...67S} {347, 67}

\bibitem[\protect\citeauthoryear{{Stern}, {Laor}  \& {Baskin}}{{Stern}
  et~al.}{2014}]{ste14}
{Stern} J.,  {Laor} A.,   {Baskin} A.,  2014, \mn@doi [\mnras]
  {10.1093/mnras/stt1843}, \href
  {https://ui.adsabs.harvard.edu/abs/2014MNRAS.438..901S} {438, 901}

\bibitem[\protect\citeauthoryear{{Tomaru}, {Done}, {Ohsuga}, {Nomura}  \&
  {Takahashi}}{{Tomaru} et~al.}{2019}]{tom19}
{Tomaru} R.,  {Done} C.,  {Ohsuga} K.,  {Nomura} M.,   {Takahashi} T.,  2019,
  \mnras, \href {https://ui.adsabs.harvard.edu/abs/2019arXiv190511763T}
  {submitted, arXiv:1905.11763}

\bibitem[\protect\citeauthoryear{{Vasudevan} \& {Fabian}}{{Vasudevan} \&
  {Fabian}}{2007}]{vas07}
{Vasudevan} R.~V.,  {Fabian} A.~C.,  2007, \mn@doi [\mnras]
  {10.1111/j.1365-2966.2007.12328.x}, \href
  {https://ui.adsabs.harvard.edu/abs/2007MNRAS.381.1235V} {381, 1235}

\bibitem[\protect\citeauthoryear{{Vasudevan} \& {Fabian}}{{Vasudevan} \&
  {Fabian}}{2009}]{vas09}
{Vasudevan} R.~V.,  {Fabian} A.~C.,  2009, \mn@doi [\mnras]
  {10.1111/j.1365-2966.2008.14108.x}, \href
  {https://ui.adsabs.harvard.edu/abs/2009MNRAS.392.1124V} {392, 1124}

\bibitem[\protect\citeauthoryear{{Wada}}{{Wada}}{2012}]{wad12}
{Wada} K.,  2012, \mn@doi [\apj] {10.1088/0004-637X/758/1/66}, \href
  {https://ui.adsabs.harvard.edu/abs/2012ApJ...758...66W} {758, 66}

\bibitem[\protect\citeauthoryear{{Wada}}{{Wada}}{2015}]{wad15}
{Wada} K.,  2015, \mn@doi [\apj] {10.1088/0004-637X/812/1/82}, \href
  {https://ui.adsabs.harvard.edu/abs/2015ApJ...812...82W} {812, 82}

\bibitem[\protect\citeauthoryear{{Woods}, {Klein}, {Castor}, {McKee}  \&
  {Bell}}{{Woods} et~al.}{1996}]{woo96}
{Woods} D.~T.,  {Klein} R.~I.,  {Castor} J.~I.,  {McKee} C.~F.,   {Bell} J.~B.,
   1996, \mn@doi [\apj] {10.1086/177101}, \href
  {http://adsabs.harvard.edu/abs/1996ApJ...461..767W} {461, 767}

\makeatother
\end{thebibliography}



\bsp	
\label{lastpage}
\end{document}